\pdfoutput=1
\documentclass[letterpaper,10pt,conference]{ieeeconf}
\IEEEoverridecommandlockouts
\usepackage[utf8]{inputenc}
\usepackage[LY1]{fontenc}
\usepackage[english]{babel}
\usepackage[babel,
  factor=450, 
  stretch=10,shrink=15, 
  tracking=true,letterspace=40 
]{microtype}
\usepackage{cite}
\usepackage{varioref}
\usepackage{balance}

\makeatletter
\let\NAT@parse\undefined
\makeatother
\usepackage[hidelinks]{hyperref}
\usepackage{booktabs}

\usepackage{tikz}
\usetikzlibrary{calc,shapes,arrows}
\usepackage{pgfplots}
\usepackage{siunitx}
\usepgfplotslibrary{units}
\makeatletter
\pgfplotsset{
  unit code/.code 2 args=
  \begingroup
  \protected@edef\x{\endgroup\si{#2}}\x
} 
\makeatother

\usepackage{sectionpatches}
\noafterindentpatch{section}
\noafterindentpatch{subsection}

\usepackage{amsmath}
\usepackage{amssymb}
\usepackage{amsbsy}
\usepackage[cmintegrals]{newtxmath}
\DeclareMathAlphabet{\mathcal}{OMS}{cmsy}{m}{n}
\usepackage{cancel}
\usepackage{mathtools}

\newtheorem{thm}{Theorem}

\def\keywords#1{\noindent{{\bfseries\slshape Index terms\,---\,}}\ignorespaces#1}

\def\trglyph{\mkern-3mu\top\mkern-4mu}
\def\tr{\ensuremath{^{\scriptstyle\trglyph}}}
\def\trs{\mkern2mu\scriptstyle\trglyph\mkern3mu}

\def\R{\ensuremath{\mathbb{R}}} 

\def\xa{\ensuremath{\mathbf{x}}}
\def\Aa{\ensuremath{\mathbf{A}}}

\def\Pa{\ensuremath{\mathbf{P}}}
\def\Qa{\ensuremath{\mathbf{Q}}}

\def\A{\ensuremath{\mathcal{A}}}

\def\P{\ensuremath{\mathcal{P}}}
\def\Q{\ensuremath{\mathcal{Q}}}
\def\J{\ensuremath{\mathcal{J}}}

\def\papertitle{State-dependent Priority Scheduling for Networked Control Systems}
\title{\papertitle}

\def\lbr{\raisebox{.07em}{\{}}
\def\rbr{\raisebox{.07em}{\}}\kern-.08em}
\def\dlm{\kern.3em\rule[-.08em]{.7pt}{1em}\kern.3em}
\author{%
  \authorblockN{%
    Ben W.\ Carabelli\authorrefmark{1},
    Rainer Blind\authorrefmark{2},
    Frank Dürr\authorrefmark{1},
    Kurt Rothermel\authorrefmark{1}%
  }\\%
  \authorblockA{%
    \parbox{.5\linewidth}{\centering%
      \authorrefmark{1}%
      Institute of Parallel and Distributed Systems\\
      University of Stuttgart, 70569 Stuttgart, Germany\\
      \lbr ben.carabelli%
      \dlm duerr%
      \dlm rothermel%
      \rbr @ipvs.uni-stuttgart.de%
    }\parbox{.5\linewidth}{\centering%
      \authorrefmark{2}%
      Institute for Systems Theory and Automatic Control\\
      University of Stuttgart, 70569 Stuttgart, Germany\\
      rainer.blind@ist.uni-stuttgart.de%
    }%
  }%
  \thanks{This work was supported by the German Research Foundation (DFG) under the research grant ``Integrated Controller Design Methods and Communication Services for Networked Control Systems (NCS)'' (RO 1086/20-1, AL 316/13-1).}
}

\def\paperkeywords{cyber-physical systems; networked control systems; optimization; scheduling}

\hypersetup{pdftitle={\papertitle}}
\hypersetup{pdfauthor={Ben W. Carabelli, Rainer Blind, Frank Dürr, Kurt Rothermel}}
\hypersetup{pdfkeywords={\paperkeywords}}

\begin{document}


\maketitle

\begin{abstract}\noindent
  Networked control systems (NCS) have attracted considerable attention in recent years.
While the stabilizability and optimal control of NCS for a given communication system has already been studied extensively, the design of the communication system for NCS has recently seen an increase in more thorough investigation.
In this paper, we address an optimal scheduling problem for a set of NCS sharing a dedicated communication channel, providing performance bounds and asymptotic stability.
We derive a suboptimal scheduling policy with dynamic state-based priorities calculated at the sensors, which are then used for stateless priority queuing in the network, making it both scalable and efficient to implement on routers or multi-layer switches.
These properties are beneficial towards leveraging existing IP networks for control, which will be a crucial factor for the proliferation of wide-area NCS applications.
By allowing for an arbitrary number of concurrent transmissions, we are able to investigate the relationship between available bandwidth, transmission rate, and delay.
To demonstrate the feasibility of our approach, we provide a proof-of-concept implementation of the priority scheduler using real networking hardware.

\end{abstract}

\begin{keywords}
  \paperkeywords
  \iftrue 
    \newlength{\bottommargin}
    \bottommargin=\paperheight
    \addtolength{\bottommargin}{-1in}
    \addtolength{\bottommargin}{-\voffset}
    \addtolength{\bottommargin}{-\topmargin}
    \addtolength{\bottommargin}{-\headheight}
    \addtolength{\bottommargin}{-\headsep}
    \addtolength{\bottommargin}{-\textheight}
    \begin{tikzpicture}[remember picture,overlay,shift={(current page.south)}]
      \node[draw,red,rectangle,thick,font=\scriptsize] at (0,.5\bottommargin) {
        \begin{minipage}{\textwidth}
          Accepted for publication at the 2017 American Control Conference (ACC), Seattle, WA, USA.
          © 2017 IEEE. Personal use of this material is permitted. Permission from IEEE must be obtained for all other uses, in any current or future media, including reprinting/republishing this material for advertising or promotional purposes, creating new collective works, for resale or redistribution to servers or lists, or reuse of any copyrighted component of this work in other works.
        \end{minipage}
      };
    \end{tikzpicture}
  \fi
\end{keywords}


\section{Introduction}\label{sec:intro}

Networked control systems (NCS) \cite{Hespanha2007} enable the economical and flexible deployment of control applications by closing feedback loops over a packet-switched network.
Applications lending themselves naturally to this kind of architecture include autonomous traffic systems, smart factories, and tele-operation, to name but a few.
With the rise of an Internet of Things (IoT) ecosystem that brings forth an increasing number of network-enabled appliances, we also expect to see more and more transient and dynamically changing NCS in the future.
All these types of applications may commonly involve widely distributed sensors, actuators and controllers.
A crucial factor for the proliferation of these applications will be to what extent existing IP networks can be leveraged by control applications.

Building NCS on top of packet-switched networks is a challenging task, because the network imposes resource constraints that conflict with the communication requirements of the control systems.
Therefore, problems relating to NCS have been studied from various angles over the last decade.
In most of these works, the communication system is assumed to be given and modelled as random packet loss or delay.
Under these assumptions, a stabilizing or optimal controller as well as bounds on the loss probability or delay are derived.
E.g., in \cite{sinopoli04a} the optimal state estimation with intermittent observations as well as bounds for the maximal allowed packet loss probability are derived.
The dual problem, i.e., optimal control over links with random packet losses is studied in \cite{imer06a,Schenato2007}.
These works are extended to more complex scenarios and more detailed models for the loss process.
E.g., in \cite{Schenato2008} the case of packet loss and delay is studied, in \cite{Garone2008} also acknowledgment packets can get lost, in \cite{Mo2013}, packet loss is modelled by a Markov chain, and in \cite{Heemels2010} trade-offs between various network related quantities are studied.
To sum up, stabilizability and optimal control for a given model of the communication system has been studied extensively.

Recently, the (co-)design of the communication system has also received increasing attention.
For instance, it has been investigated how the control performance may be improved through modifications at the transport \cite{Blind2012}, network \cite{Carabelli2014TR} or data link layer \cite{Al-Areqi2013,Al-Areqi2015,Dai2010,Molin2009,Ramesh2011,Ramesh2013,Molin2014}.
In this paper, we concern ourselves with the latter.

As already stated, most of the previous works model the network as a random loss and delay process.
Unfortunately, this does not allow strict stability guarantees, but only allows stochastic stability guarantees.
Moreover, loss probabilities and delay distributions of a network are usually neither constant nor readily available.
While such assumptions make perfect sense in wireless networks with a high degree of random channel failures, stronger guarantees are possible in switched Ethernet-based IP networks, where bandwidth restrictions and queuing delays are the main limiting factors.

In this paper, we start out under the premise that a dedicated `network slice' with fixed resources is available for a set of control systems.
This could be realized, e.g., by making a bandwidth reservation along a common route using the integrated services architecture (IntServ) \cite{RFC2212,RFC3175} of the Internet protocol stack.
Alternatively, several network virtualization technologies have been proposed \cite{Chowdhury2010} which allow provisioning of such an isolated network slice with arbitrary topologies.
Our question is then how to optimally allocate the available resources among all control systems.

This question boils down to the scheduling discipline which is used to manage the control systems' access to the network.
An early example of a dynamic scheduler for control systems is the Maximum--Error--First policy used for the Try--Once--Discard (TOD) protocol \cite{Walsh2002,Nesic2004}.
It assumes that delay-free broadcast communication is used, which may be a reasonable approximation for fieldbuses, but is an unrealistic assumption for IP networks.
In \cite{Dai2010}, a static periodic scheduling policy is derived for a set of different NCS, where the schedule is derived from average dwell times.
In \cite{Greco2011}, a stochastic RTOS scheduler for anytime control on embedded systems is studied.
In \cite{Blind2015}, the suitability of weakly hard real-time schedulers \cite{Gettings2015} has been investigated by deriving sufficient stability conditions for a single NCS, however, without directly taking network utilization and competing traffic into account.

Dynamic state-based schedulers for NCS have also been investigated.
In \cite{Molin2009,Ramesh2011}, network schedulers are designed for a single delay-free control loop, investigating under which conditions a separation principle holds for the scheduler and certainty-equivalent controller design.
This is extended in \cite{Ramesh2013,Molin2014} to multiple control loops, where each sensor uses a local scheduling policy to reduce the number of transmissions, with probabilistic contention based medium access among all NCS.
However, these works do not strictly follow a fixed resource constraint.
In \cite{Molin2014}, for instance, a price for transmissions is added to the LQR cost and adapted dynamically to maintain an upper bound on the total average transmission rate.
Therefore, while available bandwidth may be shared with other applications, a significant amount of bandwidth has to be over-provisioned in advance.

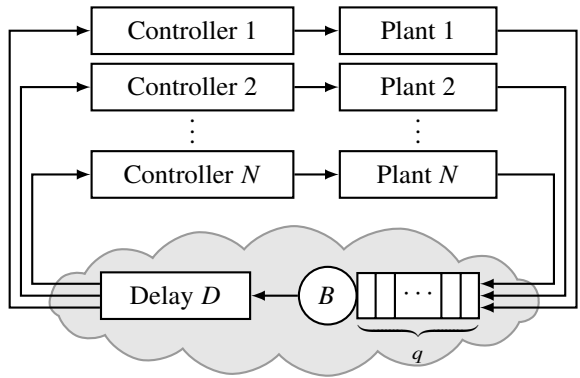
\begin{figure}
	\centering
	\vskip1em
	\begin{tikzpicture}[scale=.75, thick, >=latex,
		csnode/.style={draw,fill=white,text width=#1,text height=1.2em,text depth=0.4em,align=center}]
		\foreach \n/\y in {1/0,2/-1,N/-2.55} {
			\node[csnode=2.5cm] (c\n) at (-2,\y) {Controller $\n$};
			\node[csnode=1.9cm] (p\n) at (2,\y) {Plant $\n$};
			\draw (c\n) edge[->] (p\n);
		}
		\path (-2,-1.65) node {$\vdots$} (2,-1.65) node {$\vdots$};
		\node[draw=black!40,fill=black!10,cloud,minimum width=6.5cm,minimum height=2cm,cloud puffs=15.3,cloud puff arc=100] at (-0.2,-4.8) {};
		\node[csnode=1.8cm] (delay) at (-2.3,-4.7) {Delay $D$};
		\node[draw,fill=white,inner sep=1pt,minimum height=2.2em,rectangle split,rectangle split parts=5,rectangle split horizontal] (queue) at (2,-4.7) {\nodepart{three}$\,\cdots$};
		\node[below=-.5em] at (queue.south) {$\underbrace{\hskip1.6cm}_{q}$};
		\node[csnode=1em,circle,left=-.5pt] (server) at (queue.west) {$B\,$};
		\draw (server) edge[->] (delay);
		\foreach \n/\x/\y in {N/1/6pt,2/1.2/0,1/1.4/-6pt} {
			\draw[->] (p\n.east) -- ++(\x,0) |- ($(queue.east)+(0,\y)$);
			\draw[->] ($(delay.west)+(0,\y)$) -- ++(-\x-0.2,0) |- (c\n.west);
		}
	\end{tikzpicture}
	\caption{System architecture: $N$ different NCS communicate over a shared (virtual) link with bandwidth $B$, delay $D$, and input queue of capacity $q$.}
	\label{fig:system-model}
\end{figure}

By contrast, we propose to use priority scheduling to utilize the reserved bandwidth as well as possible to optimize control performance.
A similar approach is proposed in \cite{Al-Areqi2013,Al-Areqi2015}, where a state-based dynamic priority scheduler for physically coupled NCS is derived from a quadratically structured event-triggering rule and designed together with a suboptimal controller.
We, however, assume that the controllers are already given, and design a scheduler accordingly.
In \cite{Al-Areqi2015}, a tuning parameter is used to penalize transmissions, thereby reducing control traffic.
Like the state-based scheduling strategies mentioned in the previous paragraph, this approach assumes that only one NCS can transmit at any time.
We, on the other hand, provide formulations for an arbitrary fixed transmission rate (i.e., number of concurrent transmissions).
Also, while \cite{Al-Areqi2015} assumes that the delays are given a priori, we use a channel abstraction conforming with \cite{RFC2212} to model the relationship between available bandwidth, overall transmission rate, and delay.
We use LMI stability conditions for switched linear systems from \cite{Geromel2008} to design a scheduler that guarantees asymptotic stability and provides performance bounds for all NCS, and show how these conditions can be generalised to accommodate concurrent transmissions.
Our scheduling policy uses dynamic state-based priorities calculated at the sensors which are then used for stateless priority queuing in the network, making it both scalable and efficient to implement.
Priority queuing can be found, e.g., as part of the weighted fair queuing (WFQ) algorithm \cite{Parekh1993} supported by many routers.

Our paper is organized as follows.
In Sec.\,\ref{sec:system-model}, we present our system model, which comprises a set of NCS and the shared communication channel.
Based on this model, we formally state our scheduling problem in Sec.\,\ref{sec:problem-statement}.
We then derive a scheduler under the assumption that the communication channel admits only one transmission each sampling period in Sec.\,\ref{sec:scheduler-1}, which we then go on to generalize for an arbitrary number of transmissions per sampling period in Sec.\,\ref{sec:scheduler-q}.
In Sec.\,\ref{sec:evaluation}, we illustrate our approach with a proof-of-concept implementation and simulation examples.
A short discussion of our results and possible avenues for future work in Sec.\,\ref{sec:discussion} concludes this paper.

\section{System Model}\label{sec:system-model}

Consider a set of $N$ different NCS, each comprising a plant and controller, where state samples are sent from sensor to controller over a (virtual) network, as shown in Fig.\,\ref{fig:system-model}.
The network is shared by these NCS, but no other applications (i.e., no cross-traffic), and is modelled as a virtual link with bandwidth $B$, delay $D$, and an input queue of capacity $q$.
The queue is served by a priority scheduler.
In the following, we describe the components of the system model\,---\,control systems, virtual link, and scheduler\,---\,in more detail.

\subsection{Control System Model}

We assume that all NCS are sampled synchronously at times $t_k$ with a common sampling period $t_{k+1} - t_k = T_s$.
Each sample of a plant's state is sent in a packet addressed to the corresponding controller, together which an attached priority value.
At the same time, each controller applies a control input to the corresponding plant based on the packets that it has received so far.

The plant of NCS $i\in\{1,\dotsc,N\}$ is modelled as a discrete-time system
\begin{equation}\label{eq:plant}
	x^i_{k+1} = A^i x^i_k + B^i u^i_k,
\end{equation}
where $x^i_k\in\R^{n_i}$ is the state and $u^i_k\in\R^{m_i}$ is the input of plant $i$ at time $t_k$.
Upon sampling, the sensor sends a tuple $\bigl(x^i_k,v^i_k\bigr)$ over the network, where $v^i_k$ is the packet priority used by the scheduler to dequeue packets for transmission.
We will derive a function for calculating these priorities in Sec.~\ref{sec:scheduler-1}.
As measurements may be dropped by the scheduler due to bandwidth limitations, each controller uses the following predictive control law proposed in \cite{Blind2012}.
\begin{align}
	\hat x^i_{k+1} &= \theta(i,k) A^i x^i_k + \bigl( 1 - \theta(i,k) \bigr) A^i \hat x^i_k + B^i u^i_k, \label{eq:prediction}\\
	u^i_k &= -K^i \hat x^i_k \label{eq:controller}
\end{align}
Here, the binary arrival function $\theta(i,k)$ indicates whether the state measurement $x^i_k$ has been successfully received at the controller by time $t_{k+1}$ ($\theta=1$) or not ($\theta=0$).
The controller uses the predictive state estimate $\hat x^i_k\in\R^{n_i}$ to compensate for a constant transfer delay of $T_s$ and for dropped packets.

We assume that all plants $(A^i,B^i)$ are controllable and that a stabilizing controller $K^i$ is given, i.e., the matrices $A^i-B^iK^i$ are Schur-stable.
We use the standard LQR cost
\begin{equation}\label{eq:cost}
	J^i = \sum_{k=1}^\infty x^{i\trs}_k Q^i x^i_k + 2 x^{i\trs}_k H^i u^i_k + u^{i\trs}_k R^i u^i_k
\end{equation}
as a measure of the control performance of NCS $i$.
Therefore, a natural choice for $K^i$ is the standard infinite-horizon LQR controller, which we use in our evaluations.

However, the controller may be arbitrary, and we assume it to be designed independently from the scheduler.
Please note that we do not derive the optimal controller for this setup.
Concerning the separation of optimal control and scheduling, we refer, e.g., to \cite{Ramesh2013,Molin2014}.

\subsection{Virtual Link Model}

We assume that all NCS packets are of uniform size $L$, which accounts for the memory required for both the state and the attached priority value.
A certain bandwidth $B$ is allocated to the shared link, with an input queue of capacity $q$ (in packets) to buffer packets for forwarding.
Moreover, the link incurs a fixed delay $D$, which results from the network delay of the underlying communication service.
The end-to-end delay experienced by a batch of $q$ NCS packets is therefore $T = \frac{L}{B}q + D$.
This channel model corresponds to the end-to-end service offered, for instance, by the `Guaranteed Service' class of the IntServ architecture \cite{RFC2212} for IP networks.

Because the transfer delay of a packet must not exceed one sampling period $T_s$ in order to be useful to the controller, the queue must be dimensioned accordingly, such that
\begin{equation}\label{eq:queue-capacity}
	q \le \tfrac{B}{L}(T_s - D).
\end{equation}
Naturally, this is equivalent to a channel with a one step delay and fixed capacity $q$.
For a given (physical) network, we may trade capacity for sampling frequency subject to \eqref{eq:queue-capacity}.

As mentioned in the introduction, this shared link need not necessarily be physically restricted, but could also be realized through a virtualized network slice or an IP resource reservation, for instance.
Therefore, the bandwidth $B$ may also be regarded as a design parameter to adjust queue capacity and sampling period.

\subsection{Priority Scheduler Model}

Following the assumption of synchronous sampling, we also assume that all packets from one sampling period arrive at the queue simultaneously, as this yields the worst-case transfer delay.
The scheduler is then responsible for servicing the $q$ highest-priority packets, the remainder being dropped.
To this end, we define the set of $q$-out-of-$N$-subsets as
\begin{equation}
	\mathcal{S}_q = \Bigl\{\, S\subset\{1,\dotsc,N\} \Bigm| |S|=q \,\Bigr\} = \bigcup_{s} \bigl\{ S_s \bigr\},
\end{equation}
with an arbitrary numbering $s=1,\dotsc,{\scriptstyle\binom{N}{q}}$.
(For the simplest case $q=1$, we choose $S_s=\{s\}$.)

Now, we can define the priority scheduling function
\begin{equation}\label{eq:scheduling-function}
	\sigma(k) = \arg\min_s \sum_{i\in S_s} v^i_k,
\end{equation}
such that each NCS $i\in S_{\sigma(k)}$ receives a successful transmission in the period $[t_k,t_{k+1})$, whereas all others do not.
Thereby, the scheduling function \eqref{eq:scheduling-function} imposes a coupling on the individual arrival indicators $\theta(i,k)$ of all NCS:
\begin{equation}\label{eq:scheduler-coupling}
	\theta(i,k) = \delta\bigl(i,\sigma(k)\bigr)
	\quad\text{with}\quad
	\delta(i,s) = \begin{cases} 1 & \text{if }i\in S_s \\ 0 & \text{otherwise.} \end{cases}
\end{equation}

\section{Problem Statement}\label{sec:problem-statement}

In order to fully specify the scheduler, we need to define the priority values $v^i_k$ for the packets generated by all NCS.
The priorities must be designed such that all participating NCS remain stable under limited-capacity priority scheduling.
Within these constraints, the scheduler should optimize the overall control performance.
In order to formalize the problem statement, we consider the overall system using a lumped switching model that integrates all NCS models with the scheduling behaviour.

First, we rewrite the model of each individual NCS in a simpler form.
Plugging \eqref{eq:plant}--\eqref{eq:controller} together, we get the following autonomous model for an individual NCS with augmented state $\xa^i_k = [x^{i\trs}_k \; \hat x^{i\trs}_k]\tr$:
\begin{align}
	\xa^i_{k+1} &= \Aa^i_{\theta(i,k)} \xa^i_k,\notag
		\\
	\text{with\quad}
	\Aa^i_{\theta} &=
		\begin{bmatrix} A^i & -B^i K^i \\ \theta A^i & (1\!-\!\theta)A^i - B^i K^i \end{bmatrix}.
		\label{eq:A-individual}
\end{align}
Note that this is a switching system with two modes: $\Aa^i_0$ for open-loop and $\Aa^i_1$ for closed-loop behaviour.
The cost \eqref{eq:cost} of the individual NCS can be rewritten as
\begin{align}
	J^i &= \sum_{k=1}^\infty \xa^{i\trs}_k \Qa^i\, \xa^i_k,\notag
		\\
	\text{\quad with\quad}
	\Qa^i &=
		\begin{bmatrix} Q^i & -H^i K^i \\ -K^{i\trs} H^{i\trs} & K^{i\trs} R^i K^i \end{bmatrix}.
		\label{eq:Q-individual}
\end{align}

Next, we combine all the systems into one model for the lumped state $\eta_k = \bigl[\xa^{1\trs}_k,\xa^{2\trs}_k,\dotsc,\xa^{N\trs}_k\bigr]\tr$ of all NCS:
\begin{align}
	\eta_{k+1} &= \A_{\sigma(k)}\eta_k,
		\label{eq:lumped}\\
	\text{with\quad}
	\A_s &=
		\operatorname{diag}\Bigl(
			\Aa^1_{\delta(1,s)},\,
			\Aa^2_{\delta(2,s)},\,
			\dotsc,\,
			\Aa^N_{\delta(N,s)}
		\Bigr) \notag
\end{align}
This overall system switches between $\scriptstyle\binom{N}{q}$ modes, one for every possible outcome of the scheduler $\sigma(k)$.
The mode of each subsystem is determined by whether it is a member of the set $S_{\sigma(k)}$ of scheduled systems as determined by \eqref{eq:scheduler-coupling}.
The cost of the overall system is the sum over all NCS
\begin{align}
	\J &= \sum_{i=1}^N J^i =
		\sum_{k=1}^\infty \mathbf \eta_k\tr \Q \eta_k, \label{eq:lumped-performance}\\
	\text{with\quad}
	\Q &= 
		\operatorname{diag}\Bigl(
			\Qa^1,\,
			\Qa^2,\,
			\dotsc,\,
			\Qa^N
		\Bigr). \notag
\end{align}

Our goal is to choose the priorities $v^i_k$ which determine the scheduling function $\sigma(k)$ in \eqref{eq:scheduling-function} such that the overall system \eqref{eq:lumped} is asymptotically stable, while aiming to minimize the overall cost $\mathcal{J}$.
Moreover, the priority $v^i_k$ must depend solely on state information of the individual NCS $i$ that is available at the sensor.

\section{State-dependent Scheduler for $q=1$}\label{sec:scheduler-1}

For ease of presentation, we will begin our analysis for a link capacity of $q=1$, and later generalize the problem setting for an arbitrary queue length in Sec.~\ref{sec:scheduler-q}.
In this particular case, the scheduling function \eqref{eq:scheduling-function} simplifies to
\begin{equation*}
    \sigma(k) = \arg\min_i v^i_k.
\end{equation*}

We use the following sufficient condition from Geromel et al.~\cite{Geromel2008} to find a state-dependent, stabilizing scheduling function and performance bound for the system \eqref{eq:lumped}--\eqref{eq:lumped-performance}.
(Other stabilizing switching designs can be found, e.g., in \cite{Lin2006,Lin2009} and references therein.)

\begin{thm}[\hskip.1pt\cite{Geromel2008}]\label{thm:lyapunov-metzler}
    Let $\Q \succeq 0$ be given.
    If there exist a set of positive definite matrices $\P_1,\P_2\mkern1mu,\dotsc,\P_N$ and a matrix $\Pi$ with entries $\pi_{ij}\ge0$ and $\sum_{i=1}^N \pi_{ij} = 1,\; j=1,\dotsc,N$ satisfying the so-called Lyapunov--Metzler inequalities
    \begin{equation}\label{eq:lyapunov-metzler}
        \A_j\tr \biggl( \sum_{i=1}^N \pi_{ij} \P_i \biggr) \A_j - \P_j + \Q \prec 0
    \end{equation}
    for all $j \in \{1,\dotsc,N\}$, then the switching policy
    \begin{equation}\label{eq:centralized-scheduler}
        \sigma(k) = \arg\min_i \eta_k\tr \P_i \eta_k
    \end{equation}
    makes the origin $\eta=0$ of the system \eqref{eq:lumped} globally asymptotically stable and
    \begin{equation}\label{eq:performance-bound}
        \J =
        \sum_{k=1}^\infty \eta_k\tr \Q\, \eta_k <
        \eta_0\tr \P_{\sigma(0)} \eta_0,
    \end{equation}
    i.e., the overall performance is bounded.
    \hfill\QEDopen
\end{thm}

Note that the condition \eqref{eq:lyapunov-metzler} can also be found in the context of Markov Jump Linear Systems (MJLS).
More precisely, if $\Pi$ were the transition probability matrix of a Markov chain, \eqref{eq:lyapunov-metzler} would give mean square stability, see, e.g., \cite{Costa2005}.
However, in the considered scenario, the jumps are not driven by a Markov chain but selected deterministically by the scheduler \eqref{eq:centralized-scheduler}.
Thus, \eqref{eq:lyapunov-metzler} together with the scheduler \eqref{eq:centralized-scheduler} guarantees stability in the classical, non-stochastic sense.

Note further that finding a feasible solution to \eqref{eq:lyapunov-metzler} is a non-convex problem due to the products of $\pi_{ij}$ and $\P_i$.
However, if the matrix $\Pi$ is fixed, it becomes an LMI problem which can be solved efficiently using available numeric methods.
Therefore, we will later propose a heuristic for determining $\Pi$ using necessary stability conditions.

However, the scheduling function \eqref{eq:centralized-scheduler} depends on the full state of the lumped system, i.e.\ without further knowledge the states of all NCS have to be aggregated before a scheduling decision can be taken.
However, our system architecture requires that packet priorities only depend on the local state.

We can rectify this by imposing some constraints on the structure of the matrices $\P_i$ as follows.
For each NCS, we introduce two positive definite $2n_i\times 2n_i$ matrices $\Pa^i_0$ and $\Pa^i_1$ and add the restriction
\begin{equation}\label{eq:P-block-diagonal}
    \P_i =
    \operatorname{diag}\Bigl(
        \Pa^1_{\delta(1,i)},\,
        \Pa^2_{\delta(2,i)},\,
        \dotsc,\,
        \Pa^N_{\delta(N,i)}
    \Bigr)
\end{equation}
to the conditions in Thm.~\ref{thm:lyapunov-metzler}.
This allows us to rewrite the scheduling function \eqref{eq:centralized-scheduler} as follows:
\begin{align*}
    \sigma(k) &= \arg\min_i \eta_k\tr \P_i\, \eta_k 
               = \arg\min_i \sum_n \xa_k^{n\trs} \Pa^n_{\delta(n,i)}\, \xa_k^n \\
              &= \arg\min_i \xa_k^{i\trs} \Pa^i_1\, \xa_k^i
                 + \sum_{n\neq i} \xa_k^{n\trs} \Pa^n_0\, \xa_k^n \\
              &= \arg\min_i \xa_k^{i\trs} \bigl( \Pa^i_1 - \Pa^i_0 \bigr) \xa_k^i
                 + \cancel{\sum_n \xa_k^{n\trs} \Pa^n_0\, \xa_k^n}.
\end{align*}
As the minimum in the expression above is independent of the last term, this is equivalent to the priority scheduler \eqref{eq:scheduling-function} together with the priority functions
\begin{equation}\label{eq:priority-function}
    v^i_k = v^i(\xa^i_k) = \xa_k^{i\trs} \bigl(\Pa^i_1 - \Pa^i_0\bigr)\, \xa_k^i,
\end{equation}
which depend only on the state of the corresponding NCS.

\begin{figure*}[!t]
\normalsize
\newcounter{MYtempeqncnt}
\setcounter{MYtempeqncnt}{\value{equation}}
\setcounter{equation}{26}
\begin{align}
    \Aa_1^{i\trs} \biggl( \sum_j^{i\in S_j\mkern-10mu} \pi_{jk} \Pa^i_1 + \sum_j^{i\not\in S_j\mkern-10mu} \pi_{jk} \Pa^i_0 \biggr)\, \Aa^i_1 - \Pa^i_1 + \Qa^i &\prec 0, \quad\forall (i,k),\, i\in S_k \label{eq:LM-closed-q}\\
    \Aa_0^{i\trs} \biggl( \sum_j^{i\in S_j\mkern-10mu} \pi_{jk} \Pa^i_1 + \sum_j^{i\not\in S_j\mkern-10mu} \pi_{jk} \Pa^i_0 \biggr)\, \Aa^i_0 - \Pa^i_0 + \Qa^i &\prec 0, \quad\forall (i,k),\, i\not\in S_k \label{eq:LM-open-q}
\end{align}
\setcounter{equation}{\value{MYtempeqncnt}}
\hrulefill
\end{figure*}

In order to use this scheduler, we must first determine whether a particular set of NCS can be stabilized under this discipline, and calculate the corresponding coefficient matrices of the priority functions.
We will formulate this admission phase as an optimization problem based on the conditions in Thm.~\ref{thm:lyapunov-metzler}.
In the following, we reformulate the Lyapunov--Metzler inequalities \eqref{eq:lyapunov-metzler} and propose a heuristic for choosing the matrix $\Pi$ in order to make the problem convex.
The benefits of this formulation will become apparent in Sec.~\ref{sec:scheduler-q}, where we show how to generalize our approach to arbitrary queue lengths.

All $\P_i$ as defined in \eqref{eq:P-block-diagonal} have the same block diagonal structure as $\A_i$ and $\Q$.
Furthermore, as $q=1$, the $i^\text{th}$ diagonal block of $\A_i / \P_i$ is always given by $\Aa^i_1 / \Pa^i_1$ (closed-loop), whereas the remaining blocks are $\Aa^j_0 / \Pa^j_0$ (open-loop).
This allows us to replace the inequalities \eqref{eq:lyapunov-metzler} by the following set of lower-dimensional inequalities:
\begin{align}
    \Aa_1^{i\trs} \biggl( \pi_{ii} \Pa^i_1 + (1\!-\!\pi_{ii}) \Pa^i_0 \biggr)\, \Aa^i_1 - \Pa^i_1 + \Qa_i &\prec 0, \quad\forall_i \label{eq:LM-closed}\\
    \Aa_0^{i\trs} \biggl( \pi_{ij} \Pa^i_1 + (1\!-\!\pi_{ij}) \Pa^i_0 \biggr)\, \Aa^i_0 - \Pa^i_0 + \Qa_i &\prec 0, \quad\forall_{i\neq j} \label{eq:LM-open}
\end{align}
For each $i=1,\dotsc,N$, we can replace all inequalities in \eqref{eq:LM-open} by a single inequality by choosing $\pi_{ij}=p_i,\; \forall_{j=1,\dotsc,N}$.
For notational convenience we also define $m_i=\pi_{ii}$, which gives us
\begin{equation*}
    \Pi = \begin{bmatrix}
        m_1    & p_1    & \cdots & p_1    \\
        p_2    & m_2    & \cdots & p_2    \\
        \vdots & \vdots & \ddots & \vdots \\
        p_N    & p_N    & \cdots & m_N
    \end{bmatrix}.
\end{equation*}
If we fix all $m_i$ and $p_i$, then \eqref{eq:LM-closed}--\eqref{eq:LM-open} become LMIs in $\Pa^i_{0/1}$.
In \cite{Geromel2008}, the authors use an approach which is equivalent to choosing $m_i=\alpha\in[0,1]$ and performing a line search over $\alpha$ to accomplish this simplification.
However, since \eqref{eq:lyapunov-metzler} implies that all $\sqrt{m_i}\A_i$, $i=1,\dotsc,N$ are Schur-stable \cite{Deaecto2015}, we propose the heuristic
\begin{equation}\label{eq:ansatz-metzler-m}
    m_i=\rho(\A_i)^{-2}\cdot\alpha,
\end{equation}
where $\rho(\cdot)$ is the spectral radius, in order to exclude infeasible values a priori.
Because in Thm.~\ref{thm:lyapunov-metzler} the matrix $\Pi$ is required to be left-stochastic, i.e.\ with non-negative entries and columns summing up to 1, we can determine the coefficients $p_i$ entirely from $m_i$:
\begin{equation}\label{eq:ansatz-metzler-p-1}
    p = \bigl(\mathbf{11}\tr-I\bigr)^{-1}(\mathbf1-m),
\end{equation}
where $p$ and $m$ are the corresponding column vectors of $p_i$ and $m_i$.
This allows us to formulate our first main result.

\begin{thm}\label{thm:scheduler}
    Let $q=1$ and a set of $N$ control systems of the form \eqref{eq:plant}--\eqref{eq:cost} be given with $\theta(i,k)$ as in \eqref{eq:scheduler-coupling}.
    Let $\Aa^i_0$, $\Aa^i_1$, $\Qa^i$, $i\in\{1,\dotsc,N\}$, be defined as in \eqref{eq:A-individual}--\eqref{eq:Q-individual}.
    If there exist a scalar $\alpha\in[0,1]$ and matrices $\Pa^i_0,\Pa^i_1\succ0$ solving the semidefinite program
    \begin{align}
        \min\; & \rho \label{eq:sdp-objective}\\
        \operatorname{s.t.}\;
            & \Pa^i_1 - \rho I \prec 0, \quad\forall_i \label{eq:sdp-constraint-rho1}\\
            & \Pa^i_0 - \rho I \prec 0, \quad\forall_i \label{eq:sdp-constraint-rho0}\\
            & \Aa_1^{i\trs} \biggl( m_i \Pa^i_1 + (1\!-\!m_i) \Pa^i_0 \biggr)\, \Aa^i_1 - \Pa^i_1 + \Qa_i \prec 0, \quad\forall_i \label{eq:sdp-constraint-LM1}\\
            & \Aa_0^{i\trs} \biggl( p_i \Pa^i_1 + (1\!-\!p_i) \Pa^i_0 \biggr)\, \Aa^i_0 - \Pa^i_0 + \Qa_i \prec 0, \quad\forall_i \label{eq:sdp-constraint-LM0}
    \end{align}
    where $m$ and $p$ are defined as in \eqref{eq:ansatz-metzler-p-1} and \eqref{eq:ansatz-metzler-m},
    then the scheduler \eqref{eq:scheduling-function} with the priority functions \eqref{eq:priority-function} makes the origin of each control system globally asymptotically stable.
    Moreover, the joint LQR cost is bounded by
    \begin{equation*}
        \J<\rho\sum_{i=1}^N\|\xa^i_0\|^2.
    \end{equation*}
    \vskip-1em\hfill\QEDopen
\end{thm}

This follows from Thm.~\ref{thm:lyapunov-metzler} together with the definition of $\P_i$ in \eqref{eq:P-block-diagonal}, as shown above.
To confirm the upper bound on the control cost, we can verify that
\begin{equation*}\textstyle
    \J < \eta_0\tr \P_{\sigma(0)} \eta_0 = \sum_{i=1}^N \xa^{i\trs}_0 \Pa^i_{\sigma(0)} \xa^i_0 < \sum_{i=1}^N \rho \cdot \xa^{i\trs}_0 \xa^i_0,
\end{equation*}
due to \eqref{eq:sdp-constraint-rho1} and \eqref{eq:sdp-constraint-rho0}.

Clearly, decomposing the scheduler to enable the use of priority scheduling in the network comes at the cost of suboptimal performance compared to a centralized scheduler \eqref{eq:centralized-scheduler}, as can be seen by comparing the performance bounds in Theorems~\ref{thm:lyapunov-metzler} and \ref{thm:scheduler}.

\begin{table}
    \caption{Size of optimization problem for our approach compared to \cite{Al-Areqi2015}.}
    \label{tab:lmi-size}
    \vskip-1ex
    \centering
    \begin{tabular}{cccc}
        \toprule
        Approach & \#LMIs & size(LMI) & \#vars \\
        \midrule
        ours & $4N$ & $2n \times 2n$ & $2N(2n^2\!+\!n)+1$ \\
        \cite{Al-Areqi2015} & $>(N\!+\!1)^2$ & $4N(n\!+\!m) \times 4N(n\!+\!m)$ & $>N^3(n\!+\!m)^2$ \\
        \bottomrule
    \end{tabular}
\end{table}

In Table~\vref{tab:lmi-size} we compare the size of the optimization problem in Thm.~\ref{thm:scheduler} with that proposed in \cite{Al-Areqi2015} for a set of $N$ systems, all with state dimensions $n$ and input dimensions $m$, in terms of the number of LMI constraints, dimensions of LMI constraints, and number of scalar decision variables.
While \cite{Al-Areqi2015} co-designs a suboptimal controller in the process, our approach clearly remains more practical from a computational point of view as the number of participating NCS grows.

\section{State-dependent Scheduler for $0<q<N$}\label{sec:scheduler-q}

In the previous section, we used the special block diagonal structure of $\A_i$, $\P_i$, and $\Q$ and a heuristic for $\Pi$ to simplify the matrix inequalities in Thm.~\ref{thm:lyapunov-metzler}, and rewrite them as an optimization problem in Thm.~\ref{thm:scheduler}.
There, we made the specific assumption that $q=1$.
\addtocounter{equation}{2}%
If we drop this assumption in favor of the generalization $0<q<N$, then rewriting inequalities \eqref{eq:lyapunov-metzler} in a similar fashion yields the inequalities \eqref{eq:LM-closed-q}--\eqref{eq:LM-open-q} instead (see the top of this page).
It is easy to verify that \eqref{eq:LM-closed}--\eqref{eq:LM-open} are in a special case thereof.

As before, we can reduce this to a pair of inequalities for each NCS $i=1,\dotsc,N$.
For this purpose, the coefficients $\pi_{jk}$ of the matrix $\Pi$ must satisfy the following conditions:
\begin{align}
	\textstyle\sum_{\{j \mid\, i\in S_j\!\}} \pi_{jk} &= m_i,\quad\forall (i,k),\, i\in S_k \label{eq:ansatz-metzler-m-q}\\
	\textstyle\sum_{\{j \mid\, i\in S_j\!\}} \pi_{jk} &= p_i,\quad\forall (i,k),\, i\not\in S_k \label{eq:ansatz-metzler-p-q}
\end{align}
We assume again that the vector $m$ is given, e.g., using \eqref{eq:ansatz-metzler-m}.
As $\Pi$ now has additional degrees of freedom and its entries are not directly determined by $m$ and $p$ as they were in the previous section, we need to use a different heuristic to determine a feasible $p$.
For instance, we can use the following optimization problem:
\begin{align}
	\min_{\Pi,p}\quad & \operatorname{tr}(\Pi) \label{eq:heuristic-p:first}\\
	\operatorname{s.t.}\quad
		& \pi_{ij} \ge 0, \quad\forall_{i,j}\\
		& \!\textstyle\sum_i\, \pi_{ij} = 1, \quad\forall_j\\
		& \pi_{ii} \le \rho(\A_i)^{-2}, \quad\forall_i \label{eq:heuristic-p:schur}\\
		& \eqref{eq:ansatz-metzler-m-q}, \eqref{eq:ansatz-metzler-p-q}, \label{eq:heuristic-p:last}
\end{align}
where the first two constraints are required by Thm.~\ref{thm:lyapunov-metzler}.
The third constraint is equivalent to the necessary condition that $\sqrt{\pi_{ii}}\A_i$ must be Schur-stable, which is also our reasoning behind minimizing the trace of $\Pi$.

With \eqref{eq:ansatz-metzler-m-q}--\eqref{eq:ansatz-metzler-p-q} the inequalities \eqref{eq:LM-closed-q}--\eqref{eq:LM-open-q} become equivalent to the linear inequalities \eqref{eq:sdp-constraint-LM1}--\eqref{eq:sdp-constraint-LM0} in Thm.~\ref{thm:scheduler}.
This allows us to generalise the same result to an arbitrary queue length.

\begin{thm}\label{thm:scheduler-q}
	Let $0<q<N$ and a set of $N$ control systems of the form \eqref{eq:plant}--\eqref{eq:cost} be given with $\theta(i,k)$ as in \eqref{eq:scheduler-coupling}.
	Let $\Aa^i_0$, $\Aa^i_1$, $\Qa^i$, $i\in\{1,\dotsc,N\}$, be defined as in \eqref{eq:A-individual}--\eqref{eq:Q-individual}.
	If there exist a scalar $\alpha\in[0,1]$ and matrices $\Pa^i_0,\Pa^i_1\succ0$ solving the semidefinite program
	\eqref{eq:sdp-objective}--\eqref{eq:sdp-constraint-LM0},
	where $m$ and $p$ are defined as in \eqref{eq:ansatz-metzler-m} and \eqref{eq:heuristic-p:first}--\eqref{eq:heuristic-p:last},
	then the scheduler \eqref{eq:scheduling-function} with the priority functions \eqref{eq:priority-function} makes the origin of each control system globally asymptotically stable.
	Moreover, the joint LQR cost is bounded by $\J<\rho\sum_{i=1}^N\|\xa^i_0\|^2$.~\hfill\QEDopen
\end{thm}

Analogous to the previous section, it is straightforward to verify that $\sigma(k) = \arg\min_i \eta_k\tr \P_i\, \eta_k = \arg\min_s \sum_{i\in S_s} \xa_k^{i\trs} \bigl( \Pa^i_1 - \Pa^i_0 \bigr) \xa_k^i$.
From there, the same line of argument as for Thm.~\ref{thm:scheduler} holds.

Interestingly, this shows that the scheduling problem for any queue length can be reduced to the same semidefinite programming problem, provided the parameters $p$ are chosen appropriately according to the queue length.

Note that our previously defined scheduler requires both the plant state $x^i_k$ and the controller state $\hat x^i_k$ to be known to calculate the priorities $v^i(\xa^i_k)$.
This can be accomplished by maintaining a copy of the controller state recurrence \eqref{eq:prediction} at the sensor, for which $\theta(i,k)$ must be known, e.g., by sending acknowledgments from the controllers to the sensors, cf.\ \cite{Ramesh2013}.
However, acknowledgments would have to be delivered reliably, e.g., by reserving an additional virtual link.
Also, sending an acknowledgment of size $L_\text{ACK}$ introduces an additional delay, which must be modelled by modifying the propagation delay $D$, e.g., to $D' = 2D + \frac{L_\text{ACK}}{B}$ under the assumption of a symmetric duplex channel.

Alternatively, we can impose an appropriate structure on the optimization problem \eqref{eq:sdp-objective}--\eqref{eq:sdp-constraint-LM0}.
If we add the following constraints
\begin{equation}\label{eq:UDP-constraint}
	\Pa^i_1 = \begin{bmatrix} X^i_1 & Y^i \\[2pt] Y^{i\trs} & Z^i \end{bmatrix},\quad
	\Pa^i_0 = \begin{bmatrix} X^i_0 & Y^i \\[2pt] Y^{i\trs} & Z^i \end{bmatrix},
\end{equation}
then the priority functions become $v^i(x^i_k) = x_k^{i\trs} \bigl(X^i_1 - X^i_0\bigr)\, x_k^i$, which depend only on the plant state.
A similar approach is proposed in \cite{Geromel2008} in a different setup to design a switching output feedback controller.
Of course, the corresponding optimization problem is more conservative.
However, it allows us to achieve a higher bandwidth utilization than using acknowledgments.

\section{Evaluation}\label{sec:evaluation}

To demonstrate the feasibility of our approach, we ran experiments using real networking hardware with a set of simulated NCS and a proof-of-concept implementation of the priority scheduler.
We also show a set of pure simulation examples to illustrate the relationship between queue size, bandwidth utilization and control cost.

\subsection{Proof-of-concept Implementation}

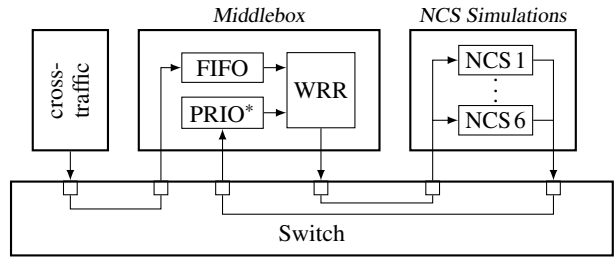
\begin{figure}
	\centering
	\begin{tikzpicture}[font=\small,>=latex,every node/.style={align=center}]
		\node[draw,thick,minimum width=8cm,text width=7cm,text height=7mm,text depth=1mm] (switch) {Switch};
		\foreach \p/\x in {1/.1,2/.25,3/.352,4/.515,5/.7,6/.9} {\path (switch.north west) -- node[anchor=north,pos=\x,draw](p\p){} (switch.north east);}
		\path (p5) -- coordinate[pos=.5](pncs) (p6);
		\draw (p1) -- ++(0,-.28) -| (p2);
		\draw (p3) -- ++(0,-.35) -| (p6);
		\draw (p4) -- ++(0,-.2) -| (p5);
		\node[draw,thick,yshift=1.3cm,rotate=90,minimum width=1.6cm,minimum height=1cm] (trafgen) at (p1) {cross-\\traffic};
		\node[draw,thick,yshift=1.3cm,minimum height=1.6cm,minimum width=2.2cm] (ncsnode) at (pncs) {};
			\node[above,font=\footnotesize\slshape] at (ncsnode.north) {NCS Simulations};
		\path (trafgen) --node[pos=.5,draw,thick,minimum height=1.6cm,minimum width=3.2cm](middlebox){} (ncsnode);
			\node[above,font=\footnotesize\slshape] at (middlebox.north) {Middlebox};

		\node[draw,yshift=1.6cm,text width=9mm] (fifo) at (p3) {FIFO};
		\node[draw,yshift=1.0cm,text width=9mm] (prio) at (p3) {PRIO$^*$};
		\node[draw,yshift=1.3cm,minimum height=1cm] (wrr) at (p4) {WRR};
		\node[draw,xshift=1pt,yshift=1.7cm,text width=8mm] (ncs1) at (pncs) {NCS\,1};
		\node[draw,xshift=1pt,yshift=0.9cm,text width=8mm] (ncs6) at (pncs) {NCS\,6};
		\path (ncs1) -- node[pos=.25]{$\vdots$} (ncs6);

		\draw[->] (trafgen) -- (p1);
		\draw[->] (p2) |- (fifo);
		\draw[->] (p3) -- (prio);
		\draw[->] (fifo) -- (fifo-|wrr.west);
		\draw[->] (prio) -- (prio-|wrr.west);
		\draw[->] (wrr) -- (p4);
		\draw[->] (p5) |- coordinate (branch-left) (ncs6);
		\draw[->] (branch-left) |- (ncs1);
		\draw[->] (ncs6) -| coordinate (branch-right) (p6);
		\draw (ncs1) -| (branch-right);
	\end{tikzpicture}
	\caption{%
		Networking testbed setup used for evaluation.
		$^*$) Priority scheduling and deficit round robin scheduling were implemented for comparison.
	}
	\label{fig:eval-setup}
\end{figure}

We ran a proof-of-concept evaluation in a networking testbed consisting of commodity machines (Intel Xeon E5-1650) with $4\times10$\,Gbps Ethernet network interface cards, connected to a 10\,Gbps Ethernet switch (Edge-Core AS5712-54X).
The setup of our testbed is shown in Fig.~\ref{fig:eval-setup}.

A set of $N=6$ NCS were simulated on one of the machines, with outgoing measurement packets from the simulated sensors of all NCS instances being sent over one network interface, and incoming packets being received on a different interface.
In order to improve throughput and reduce unpredictable delays, we used Intel's Data Plane Development Kit (DPDK) \cite{DPDK} instead of Sockets for sending and receiving packets from the simulation instances.
We used identical models of an inverted pendulum on a cart for all 6 NCS simulation instances (for details see \cite{Zinkler2016}).
The sampling period was chosen at $T_s=50$\,ms.
The pendulum for NCS 1 was started at an initial angle of $\phi=35\si\degree$, while all others were started at the origin $\phi=0\si\degree$.

A second machine was dedicated to producing cross-traffic approximately at line rate on a dedicated network interface.
We use this cross-traffic to demonstrate that our proof-of-concept implementation realizes a dedicated virtual link to isolate NCS traffic from the effects of traffic from other applications, which is one of our initial assumptions.

A third machine was used to host a software middlebox implementation realizing both the virtual link provisioning and priority scheduling, also using DPDK for networking.
The middlebox maintains two input queues: a priority queue with capacity $q=2$ for NCS traffic, and a FIFO queue for all other traffic.
The two queues are served in a weighted round robin fashion, where the priority queue is fully served every $T_s=50$\,ms and the FIFO queue is served during the remaining time.
All outgoing traffic is transmitted on the same interface, but cross-traffic is discarded at the switch while NCS traffic is forwarded to the NCS simulation node.
In order to compare our state-based dynamic scheduler to a fair static scheduler, we also implemented an alternative middlebox, with the difference that the queue for NCS traffic is not served using packet priorities, but using a deficit round robin (DRR) protocol.

\begin{figure}
	\centering
	\begin{tikzpicture} \pgfplotsset{footnotesize}
		\begin{axis}[
			name=RoundRobin,legend cell align=left,
			width=7cm,height=2cm,scale only axis,grid=major,
			xticklabels={},xmin=0,xmax=300,xtick={0,50,...,300},
			ylabel=$\phi\!$,y unit=\si{\degree},ymin=-99,ymax=99,ylabel near ticks,y label style={rotate=-90,xshift=1ex},ytick={-90,-45,0,45,90}
		]
			\addplot[black,thick]       	table[x index=0,y index=1] {eval/ncs_coord_rr.txt};
			\addplot[black!50,very thin]	table[x index=0,y index=2] {eval/ncs_coord_rr.txt};
			\addplot[black!50,very thin]	table[x index=0,y index=3] {eval/ncs_coord_rr.txt};
			\addplot[black!50,very thin]	table[x index=0,y index=4] {eval/ncs_coord_rr.txt};
			\addplot[black!50,very thin]	table[x index=0,y index=5] {eval/ncs_coord_rr.txt};
			\addplot[black!50,very thin]	table[x index=0,y index=6] {eval/ncs_coord_rr.txt};
			\addplot[black,thick]       	table[x index=0,y index=1] {eval/ncs_coord_rr.txt};
			\addplot[mark=none, black, dashed] coordinates {(-2,90) (300,90)};
			\addplot[mark=none, black, dashed] coordinates {(-2,-90) (300,-90)};
			\legend{NCS 1,NCS 2--6}
		\end{axis}
		\begin{axis}[
			name=PriorityScheduling,legend cell align=left,
			at={(RoundRobin.below south west)},anchor=north west,yshift=2em,
			width=7cm,height=2cm,scale only axis,grid=major,
			xlabel=$t$,xmin=0,xmax=300,x unit=\si{\second},xlabel near ticks,xtick={0,50,...,300},
			ylabel=$\phi\!$,y unit=\si{\degree},ymin=-99,ymax=99,ylabel near ticks,y label style={rotate=-90,xshift=1ex},ytick={-90,-45,0,45,90}
		]
			\addplot[black,thick]       	table[x index=0,y index=1] {eval/ncs_coord_p.txt};
			\addplot[black!50,very thin]	table[x index=0,y index=2] {eval/ncs_coord_p.txt};
			\addplot[black!50,very thin]	table[x index=0,y index=3] {eval/ncs_coord_p.txt};
			\addplot[black!50,very thin]	table[x index=0,y index=4] {eval/ncs_coord_p.txt};
			\addplot[black!50,very thin]	table[x index=0,y index=5] {eval/ncs_coord_p.txt};
			\addplot[black!50,very thin]	table[x index=0,y index=6] {eval/ncs_coord_p.txt};
			\addplot[black,thick]       	table[x index=0,y index=1] {eval/ncs_coord_p.txt};
			\addplot[mark=none, black, dashed] coordinates {(-2,90) (300,90)};
			\addplot[mark=none, black, dashed] coordinates {(-2,-90) (300,-90)};
			\legend{NCS 1,NCS 2--6}
		\end{axis}
	\end{tikzpicture}
	\vskip-1ex
	\caption{%
		Time series of pendulum angles from evaluation of proof-of-concept implementation with $N=6$, $q=2$, and $T_s=50$\,ms;
		top: round robin scheduling;
		bottom: state-based priority scheduling.
	}
	\label{fig:eval-timeseries}
\end{figure}
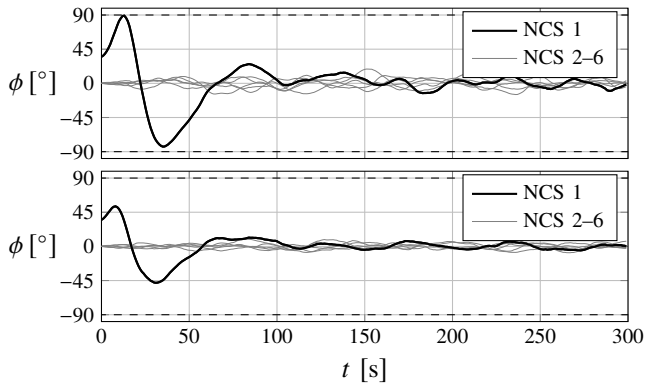

\begin{table}
    \caption{%
    	LQR cost comparison for proof-of-concept implementation
    }
    \label{tab:eval-cost}
    \vskip-1ex
    \centering
    \begin{tabular}{cccccccc}
        \toprule
        Scheduler & $\J$ & $J^1$ & $J^2$ & $J^3$ & $J^4$ & $J^5$ & $J^6$ \\
        \midrule
        Round-robin   & 73.0 & 48.0 & 5.3 & 4.3 & 3.8 & 7.3 & 4.3 \\
        Priority & 23.0 & 13.7 & 2.1 & 1.8 & 1.8 & 1.5 & 2.1 \\
        \bottomrule
    \end{tabular}
\end{table}

We ran evaluations for both configurations: once using naive deficit round robin and once using our state-based priority scheduling for NCS traffic.
Fig.~\ref{fig:eval-timeseries} shows time series for the angle $\phi$ of all six simulated pendulums, while Table~\ref{tab:eval-cost} shows joint and individual LQR cost for comparison.
Using state-based priority scheduling reduces the joint cost $\J$ by 68\% compared to round robin.
Moreover, the cost is more evenly distributed between the NCS.
While the stretch between the worst and best performing NCS is $\frac{\max_iJ^i}{\min_iJ^i}=12.8$ with round robin, it is only $8.8$ with state-based priority scheduling.

\subsection{Simulation Example}

To illustrate the effects of queue dimensioning, we show some simulation results for a fixed set of NCS and network model with varying $q$.
We simulate a link with bandwidth $B=10\,000\,\frac{\mathrm{bit}}{\mathrm{s}}$ and delay $D=20\,\mathrm{ms}$, that is shared by $N=10$ NCS with identical (continuous) plant dynamics
\begin{equation*}
	\frac{\operatorname{d}}{\operatorname{d}t} x^i(t) =
	    \begin{bmatrix} 0 & 1 \\ -2 & 2 \end{bmatrix} x^i(t) +
	    \begin{bmatrix} 0 \\ 1 \end{bmatrix} u^i(t) + w^i(t),
\end{equation*}
initial conditions $x^i(t_0)=\bigl[1,1\bigr]\tr$, LQR controller for $Q=I$ and $R=0.1$, and additive white noise $w^i(t)\sim\mathcal{N}\bigl(0,10^{-3}I\bigr)$.
We assume the packet size to be $L=192\,\mathrm{bit}$, which corresponds to two IEEE double-precision floating point values for the state and one for the priority.
The overall system with the state-based scheduler is simulated for $q=1,\dotsc,N$ over a time period of 100 seconds.
In each simulation, the systems are discretized to the minimum possible sampling time according to the link model, i.e.\ $T_s = \frac{L}{B}q + D$.

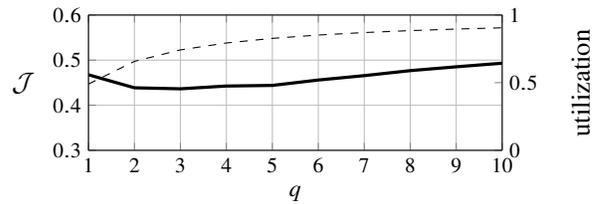
\begin{figure}
	\centering
	\begin{tikzpicture}
		\pgfplotsset{footnotesize,set layers}
		\begin{axis}[width=5.5cm,height=1.8cm,scale only axis,grid=major,
			xlabel=$q$,xmin=1,xmax=10,xlabel style={yshift=1ex},
			axis y line*=left,ylabel near ticks,ylabel style={rotate=-90,xshift=-0.2em},
			ylabel=$\J$,ymin=0.3,ymax=0.6]
			\addplot[black,very thick] coordinates {
				(1,0.467538)
				(2,0.438682)
				(3,0.436419)
				(4,0.442590)
				(5,0.443952)
				(6,0.455901)
				(7,0.465654)
				(8,0.476727)
				(9,0.485225)
				(10,0.493196)
			};
		\end{axis}
		\begin{axis}[width=5.5cm,height=1.8cm,scale only axis,
			axis x line=none,xmin=1,xmax=10,
			axis y line*=right,ylabel near ticks,
			ylabel=utilization,ymin=0,ymax=1,ytick={0,0.5,1},ylabel style={yshift=-1em}]
			\addplot[black,dashed] coordinates {
				(1,0.489796)
				(2,0.657534)
				(3,0.742268)
				(4,0.793388)
				(5,0.827586)
				(6,0.852071)
				(7,0.870466)
				(8,0.884793)
				(9,0.896266)
				(10,0.90566)
			};
		\end{axis}
	\end{tikzpicture}
	\vskip-1ex
	\caption{
		Simulations for varying queue capacity $q$:
		joint cost $\J$ (thick) and bandwidth utilization (dashed).
	}
	\label{fig:example}
\end{figure}

The results, averaged over 50 realizations of the simulation, can be seen in Fig.~\ref{fig:example}.
The plot shows the joint LQR cost $\J$, together with the bandwidth utilization, which is given by the ratio of the transmission rate $\frac{qL}{T_s}$ to the available bandwidth $B$.
Note that the case $q=10$ corresponds to a static equal-bandwidth schedule.
We can see that the cost decreases with decreasing $q$, corresponding to more dynamic scheduling, but increases again towards $q=1$.
This can be attributed to a lower bandwidth utilization, as the propagation delay $D$ dominates the sampling period for small $q$.

\section{Summary and Discussion}\label{sec:discussion}

In this paper, we addressed the optimal scheduling problem for a set of NCS sharing a dedicated network slice.
We introduced a switched model of the overall system with a limited-capacity queue model for the communication channel.
Based on LMI stability conditions for switched linear systems from \cite{Geromel2008}, we first designed a state-based priority scheduler for a channel capacity of one transmission per sampling period.
We then generalised our scheduler design to allow an arbitrary number of NCS to transmit concurrently within one sampling period.
The resulting scheduling policy guarantees performance and asymptotic stability of all NCS, and only requires stateless priority queuing in the network, making it both scalable and efficient to implement.

To conclude this work, we would like to discuss some aspects of our results.
We designed a scheduler under the assumption that all NCS come with a given controller.
Of course, joint optimal design of scheduling and controller, e.g.\ as in \cite{Al-Areqi2015}, is an interesting topic for future work.
Furthermore, it could be studied how to choose the queue length for a given available bandwidth and network delay, such as to optimize the overall control performance.
Also, while we considered priority scheduling for full state feedback here, the output feedback case can be studied by adding restrictions to the LMI constraints, as proposed in \cite{Geromel2008}.

Our approach does not provide isolation between the traffic of individual NCS.
While this allows us to utilize the available bandwidth to improve overall performance, it opens up the opportunity for one or more NCS to use more of their ``fair'' share to the detriment of all others, possibly to the point of instability.
Apart from malicious priority inflation, modelling errors could also lead to this kind of behaviour.
One possible solution is to use traffic shaping to limit the bandwidth available to each NCS.
The consideration of additional shaping constraints is a topic for future work.

In our system model, we make some assumptions that should also be discussed briefly.
First, sampling times and therefore packet arrival times of all NCS are synchronized.
If this assumption is violated, i.e.\ sampling times of different NCS are phase shifted or packets experience different delays between sensor and scheduler, then any low-priority packet arriving early could impose an additional queuing delay on the remaining NCS traffic and might ultimately cause a higher-priority packet to be dropped.
Moreover, we did not account for priorities from different (overlapping) sampling periods to be compared at the scheduler.

Second, the communication channels of all NCS are modelled by one shared link.
However, in a realistic application scenario, it should be assumed that the traffic of different NCS is routed over overlapping multi-hop paths.
This means that a scheduling decision may be required at every hop.
In principle, the same scheduler could be implemented at all network elements, since the priorities are preserved under scheduling of different packet subsets in an arbitrary order.
However, even if the sampling times are synchronized, this is not necessarily true for the arrival times at schedulers in the network, which has been discussed above.
Also, it may lead to suboptimal resource utilization, since unnecessary constraints may be imposed on traffic flows with mutually disjoint network paths.
How our approach can be extended to address these issues is to be investigated in future work.

\section*{Acknowledgments}

The authors would like to thank the German Research Foundation (DFG) for financial support, and the anonymous reviewers for their thoughtfulness and helpful suggestions.


\end{document}